\documentclass[aps,prl,twocolumn,groupedaddress,footinbib]{revtex4-2}

\usepackage{amsmath}

\usepackage[dvips]{color}

\usepackage[latin2]{inputenc}

\usepackage[T1]{fontenc}

\usepackage[american]{babel}

\usepackage{color}

\usepackage{graphicx}

\usepackage{epstopdf}

\usepackage{epsfig}

\definecolor{nred} {RGB}{224,0,0}

\newcommand{\IM}{\mbox {$Im$}}

\allowdisplaybreaks

\begin{document}

\title{Exact solution of electronic transport in semiconductors dominated by scattering on polaronic impurities}

\author{J. Krsnik$^{1}$} 
\email{jkrsnik@ifs.hr} 
\author{I. Batisti\' c$^2$, A. Marunovi\'c$^3$, E. Tuti\v s$^1$, and O. S. Bari\v si\' c$^1$}

\affiliation{$^{1}$Institute of Physics, Bijeni\v cka c. 46, HR-10000 Zagreb, Croatia,\\
$^{2}$Department  of  Physics,  Faculty  of  Science, University  of  Zagreb,  HR-10000  Zagreb,  Croatia\\
$^{3}$Shell Global, Carel van Bylandtlaan 16, The Hague, The Netherlands}

\begin{abstract}
The scattering of electrons on impurities with internal degrees of freedom is bound to produce the signatures of the scatterer's own dynamics and results in nontrivial electronic transport properties.  Previous studies of polaronic impurities in low-dimensional structures, like molecular junctions and one-dimensional nanowire models, have shown that perturbative treatments cannot account for a complex energy dependence of the scattering cross section in such systems.  Here we derive the exact solution of polaronic impurities shaping the electronic transport in bulk (3D) systems. In the model with a short-ranged electron-phonon interaction, we solve for and sum over all elastic and inelastic partial cross sections, abundant in resonant features.  The temperature dependence of the charge mobility shows the power-law dependence, $\mu(T)\propto T^{-\nu}$, with $\nu$ being highly sensitive to impurity parameters. The latter may explain nonuniversal power-law exponents observed experimentally, e.g. in high-quality organic molecular semiconductors.
\end{abstract}

\maketitle

Effects of the electron-phonon interaction (EPI) in semiconductors \cite{Devreese} are frequently analyzed in the context of polaron \cite{Landau} formation. Namely, the itinerant charge couples to crystalline phonons, moving as a dressed quasi-particle along the crystal lattice \cite{BB}. In addition to these delocalized states, localized polaron states may form when impurities are introduced into the system \cite{Hague,Mishchenko,Ebrahimnejad}. Properties of the latter have been successfully investigated by electron spin resonance, which directly reveal microscopic details of the EPI, in semiconductor crystals and thin-film transistors \cite{Matsui,Seki}. 

Regarding transport properties, the polaronic coupling to a phonon degree of freedom at the impurity site has attracted great attention in the context of single-electron tunneling across microscopic junctions, nanowires, and quantum dots \cite{Zhitenev,Zimbovskayaa}. Steps in the I-V characteristic curves \cite{Dong}, phonon-assisted tunneling \cite{Cai}, effects of Franck-Condon \cite{Leturcq}, Coulomb \cite{Park}, and bipolaron \cite{Fang} blockade have been reported both experimentally and theoretically \cite{Wingreen}.  Surprisingly, given the large number of interesting features found in 1D systems, the role of polaronic impurities in systems with $D>1$, (e.g. $D=3$) has not been investigated so far, to the best of our knowledge. Therefore, our study concentrates on the question of how the presence of polaronic impurities affects the mobility of electrons. Indeed, in semiconductors one may easily imagine such impurities as a strong source of scattering for charge carriers. In particular, we find that the scattering on polaronic impurities may explain transport properties observed in some transition-metal oxide systems \cite{Bosman} and organic semiconductors \cite{Karl,Xi,Krupskaya,Podzorov}. Actually, very recently \cite{Mettan}, these impurities have been identified as strong phonon scatterers, responsible for a drastic suppression of the thermal conductivity in anatase TiO$_2$ single-crystals.

{\it Polaronic impurity problem.} For a low concentration $n_{i}$ of randomly distributed impurities, correlations between scattering events involving multiple  different impurities may usually be neglected. The scattering rate is then proportional to $ n_{i}\sigma$, where $\sigma$ is the single-impurity cross-section. In this context, we analyze the single-impurity model that involves a coupling between the electron and the local lattice deformation at the impurity site, in addition to a change of the electron orbital energy. In the standard notation for the electron ($c^\dagger,c$) and phonon ($a^\dagger,a$) operators the Hamiltonian is given by,

\begin{equation} 
\hat H =\sum_\textbf{k}\varepsilon_\textbf{k}c^\dagger_\textbf{k}c_\textbf{k} + \omega_0\;a_\textbf{l}^\dagger a_\textbf{l}
+ \left[ \varepsilon_0 + g (a_\textbf{l}^\dagger+a_\textbf{l})\right] c^\dagger_\textbf{l}c_\textbf{l} \;.\label{Ham}
\end{equation}

\noindent Here, $\textbf{k}$ denotes the electron wave vector, $c^\dagger_\textbf{k}=\sum_{\textbf{j}}e^{i\textbf{k}\textbf{j}} c^\dagger_\textbf{j}$, whereas $\textbf{j}=\textbf{l}$ denotes the impurity site that breaks the translational symmetry of the lattice. The model~(\ref{Ham}) allows for an arbitrary system dimension and an electron dispersion $\varepsilon_\textbf{k}$, while the (neutral) polaronic impurity is modeled by three parameters, the orbital energy $\varepsilon_0$, the phonon energy $\omega_0$ ($\hbar=1$), and the strength of the short-range (Holstein) EPI $g$. For the rest of the system, we assume that the EPI plays a minor role in scattering processes in comparison to effects caused by the scattering on polaronic impurities.

While the exact results of the single-electron problem (\ref{Ham}) have been obtained numerically for the 1D case \cite{Haule}, here we derive its exact solution in the closed-form for an arbitrary dimension. The 3D case is in our focus in the present paper, whereas the application of the same approach to 1D is provided for comparison in Supplemental Materials \cite{Suppl}. We use  the quantum field-theoretical technique that has its direct interpretation in terms of Feynman diagrams. In this sense, our expressions are most general and may be applied to any system with known unperturbed electron and phonon propagators, including problems with impurities near surfaces.  As in exact treatments of 1D systems, we do not take into the account a scenario when the electron finds the polaronic impurity site being occupied by another electron. Indeed, for small impurity concentrations or shallow localized states like in rubrene \cite{Krellner}, this situation is limited to low temperatures only, similarly to ordinary semiconductors operating in the extrinsic ionization regime~\cite{Sapoval}. In fact, we find very different scattering regimes even when all impurity states lie above the bottom of the conduction band.

{\it Exact solution.} In order to treat in a unifying manner electron scattering processes that preserve (elastic) or change (inelastic) the number of phonons in the system, we consider a generalized unperturbed Green function (GF) operator $\hat G^{(0)}$ \cite{Goodvin,cini,Hahn}, which matrix elements in the real-space representation are given by,

\begin{equation}  \label{G0}
G^{(0)\gamma,\alpha}_{\textbf{n},\textbf{m}}(\omega) = \left\langle 0\right| \frac{\left( a_{\textbf{l}}\right) ^{\gamma}}{\sqrt{\gamma!}}c_{\textbf{n}}\frac{1}{\omega -\hat H_{0}+i\eta} c^{\dagger}_{\textbf{m}}\frac{(a_{\textbf{l}}^{\dagger})^{\alpha}}{\sqrt{\alpha!}}\left| 0\right\rangle\;.
\end{equation}

\noindent Here, $\hat H_0$ corresponds to the first two terms in Eq.~(\ref{Ham}), involving electron and phonon degrees of freedom. The exact GF operator $\hat G$ is obtained by taking the full Hamiltonian instead of $\hat H_0$ in Eq.~(\ref{G0}). $\hat G^{(0)}$  is diagonal in the number of initial $\alpha$ and final $\gamma$ phonons, $\gamma=\alpha$, whereas $\hat G$  involves transitions between different phonon states due to the EPI, $V^{\gamma,\alpha}_{\textbf{l},\textbf{l}}=\varepsilon_0\delta_{\gamma,\alpha}+g(\sqrt{\gamma}\delta_{\gamma+1,\alpha}+\sqrt{\alpha}\delta_{\gamma,\alpha-1})$, given by the third term in Eq.~(\ref{Ham}). With $\hat V$ involving the impurity site only, the matrix elements of $\hat G$ satisfy,

\begin{equation} \label{RRSE}
G^{\gamma,\alpha}_{\textbf{n},\textbf{m}} = \delta_{\gamma,\alpha}G^{(0)\alpha,\alpha}_{\textbf{n},\textbf{m}} + G^{(0)\gamma,\gamma}_{\textbf{n},\textbf{l}}\sum_\zeta V^{\gamma,\zeta}_{\textbf{l},\textbf{l}}G^{\zeta,\alpha}_{\textbf{l},\textbf{m}}\;.
\end{equation}

\noindent As shown in Supplemental Material \cite{Suppl}, it is possible to rewrite Eq.~(\ref{RRSE}) in terms of an operator that acts at the impurity site only, $\Gamma^{\gamma,\alpha}_{\textbf{n},\textbf{m}}=\delta_{\textbf{n},\textbf{l}}\delta_{\textbf{m},\textbf{l}}\Gamma^{\gamma,\alpha}$, as,

\begin{equation} 
G^{\gamma,\alpha}_{\textbf{n},\textbf{m}} = 
\delta_{\gamma,\alpha} G^{(0)\alpha,\alpha}_{\textbf{n},\textbf{m}}+G^{(0)\gamma,\gamma}_{\textbf{n},\textbf{l}}\Gamma^{\gamma,\alpha}
G^{\alpha,\alpha}_{\textbf{l},\textbf{m}}\;,\label{Gabnm}
\end{equation}

\noindent which for the elastic part of the problem at the impurity site gives rise to the Dyson form, $[G^{\alpha,\alpha}_{\textbf{l},\textbf{l}}]^{-1}=[G^{(0)\alpha,\alpha}_{\textbf{l},\textbf{l}}]^{-1}-\Gamma^{\alpha,\alpha}$. Combination of this Dyson form and Eq.~(\ref{Gabnm}) yields,

\begin{equation} \label{TM}
G^{\gamma,\alpha}_{\textbf{n},\textbf{m}} = \delta_{\gamma,\alpha}G^{(0)\alpha,\alpha}_{\textbf{n},\textbf{m}} + 
G^{(0)\gamma,\gamma}_{\textbf{n},\textbf{l}}\frac{\Gamma^{\gamma,\alpha}}{1-G^{(0)\alpha,\alpha}_{\textbf{l},\textbf{l}}
	\Gamma^{\alpha,\alpha}}G^{(0)\alpha,\alpha}_{\textbf{l},\textbf{m}}\;,
\end{equation}

\noindent where, as discussed below, the exact $\Gamma^{\gamma,\alpha}$ may be found in a closed-form.

\begin{figure}[t]
	\includegraphics[width=0.5\columnwidth]{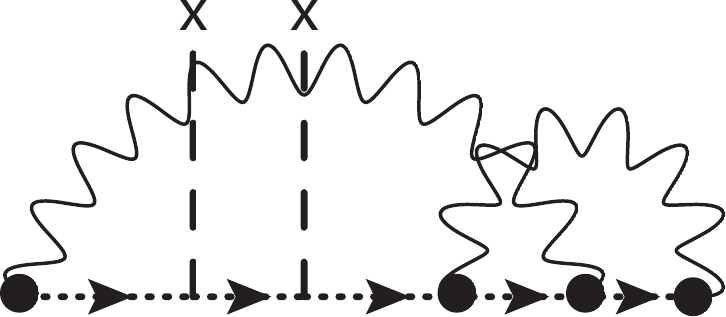}
	\caption{$\Gamma^{\alpha,\alpha}$ diagram for $\alpha=0$, involving the static $\varepsilon_0\neq0$ scattering (vertical dashed lines) and the dynamic scattering on phonons (wavy lines). Dotted lines represent the unperturbed electron propagator.}
	\label{fig01}
\end{figure}

A diagrammatic representation of the exact solution provides valuable insights on the frequency characteristics of relevant scattering events. A typical diagram corresponding to $\Gamma^{\alpha,\alpha}$ is shown in Fig.~\ref{fig01} for $\alpha=0$, when $G^{(0)0,0}_{\textbf{l},\textbf{l}}$ is the unperturbed local electron propagator and $\Gamma^{0,0}$ is the electron self-energy. In Fig.~\ref{fig01}, $G^{(0)0,0}_{\textbf{l},\textbf{l}}$ is represented by the horizontal dotted lines, the vertical dashed lines correspond to the static $\varepsilon_0\neq 0$ scattering, whereas the wavy lines correspond to the phonon propagators. For the current single-electron problem there is no renormalization of phonon lines.

On the diagrammatic level, it is easy to see that the static scattering may be summed up separately from the dynamic EPI contribution. Namely, for $g=0$, the effect of finite $\varepsilon_0$ is easily accounted for in the exact manner, since $G^{\alpha, \alpha}_{\textbf{l},\textbf{l}}(\omega)|_{g=0}= [G^{(0)\alpha,\alpha}_{\textbf{l},\textbf{l}}(\omega)]^{-1}-\varepsilon_0$. Therefore, hereafter we assume that the effect of finite $\varepsilon_0$ is included in the $g=0$ impurity propagator, $G_I(\omega-\alpha\omega_0)=G^{\alpha, \alpha}_{\textbf{l},\textbf{l}}(\omega)|_{g=0}$. This approach may be generalized to any distribution of static impurities (including changes of hopping integrals) since this does not affect the structure of the diagrammatic expansion in $g$. In particular, the diagonal matrix elements, $\Gamma^{\alpha,\alpha}$, giving rise to the elastic scattering, may be expressed in the continued fraction form \cite{cini,cini2}, $\Gamma^{\alpha,\alpha} = g\alpha A_\alpha+gB_\alpha$, with $A_\alpha$ and $B_\alpha$ representing processes with phonon absorption and emission respectively,

\begin{eqnarray} 
A_{\alpha}(\omega) &=& \frac{g}{G^{-1}_I(\omega - (\alpha -1) \omega_{0})-\frac{(\alpha-1)g^{2}}{G_I^{-1}(\omega - (\alpha-2)\omega_{0})-...}}\nonumber\\
B_{\alpha}(\omega) &=& \frac{(\alpha+1)g}{G_I^{-1}(\omega - (\alpha+1)\omega_{0})-\frac{(\alpha+2)g^{2}}{G_I^{-1}(\omega - (\alpha+2)\omega_{0})-...}}\;.\label{AB}
\end{eqnarray}

\noindent The inelastic contributions are given by $\Gamma^{\gamma,\alpha}$ \cite{Suppl},

\begin{equation} \label{Sab}
\Gamma^{\gamma,\alpha}= \begin{cases}
g\sqrt{\frac{\alpha!}{\gamma!}}\left( \gamma + B_{\gamma}B_{\gamma-1}\right)\prod_{i=\alpha}^{\gamma-2}B_{i}, &\gamma>\alpha + 1,\\
g\sqrt{\frac{\alpha!}{\gamma!}}\left( \gamma + B_{\gamma}B_{\alpha}\right), &\gamma=\alpha + 1,\\
g\sqrt{\frac{\alpha!}{\gamma!}}\left( 1 + \gamma A_{\gamma}A_{\alpha}\right), &\gamma=\alpha - 1,\\
g\sqrt{\frac{\alpha!}{\gamma!}}\left( 1 + \gamma A_{\gamma}A_{\gamma+1}\right)\prod_{i=\gamma+2}^{\alpha}A_{i}, &\gamma<\alpha - 1\;.
\end{cases}
\end{equation}

\noindent By inspecting the continued fraction expansion order by order in $g$, these equations may be put in direct correspondence with the Feynman diagrams. For example, up to $g^3$, for the inelastic processes with no phonons in the initial and one phonon in the final state, one obtains,

\begin{equation} 
\Gamma^{1,0}= g(1+B_{1}B_{0})\approx g+
2g^3G_I(\omega-\omega_0)
G_I(\omega-2\omega_0)\;.\label{S10g3}
\end{equation}

\begin{figure}[t]
	\includegraphics[width=0.8\columnwidth]{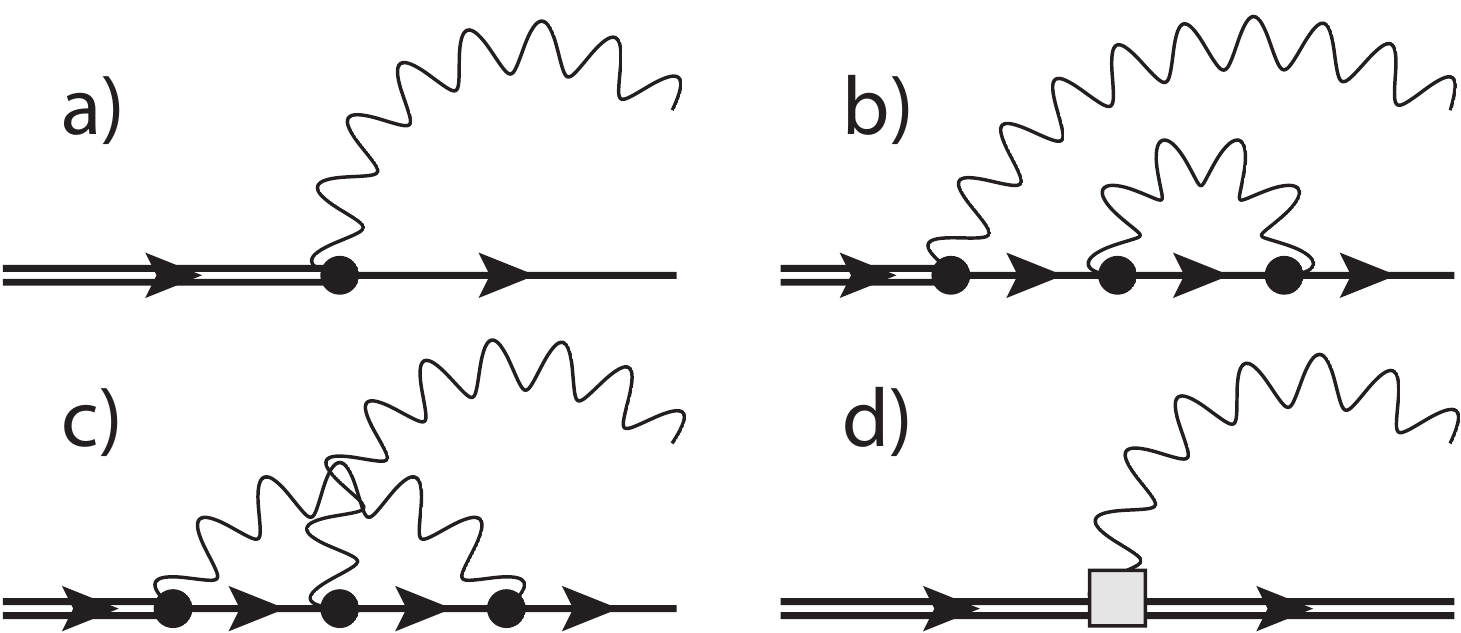}
	\caption{GF diagrams. Single lines represent the electron propagator with static scattering included, double lines represent the exact one. The square in the fourth diagram is the exact electron-phonon vertex function.}
	\label{fig02}
\end{figure}

\noindent In the expansion of the GF, the first term in Eq.~(\ref{S10g3}) corresponds to Fig.~\ref{fig02}a, while the second corresponds to the two diagrams shown in Fig.~\ref{fig02}b and Fig.~\ref{fig02}c, with equal contributions. In particular, Fig.~\ref{fig02}b shows the leading correction of the outgoing electron propagator, while Fig.~\ref{fig02}c shows the leading vertex correction of the phonon emission process. As shown in Fig.~\ref{fig02}d, in the infinite order in $g$, $\Gamma^{1,0}$ involves all the corrections of the outgoing electron propagator and all the vertex corrections of the phonon emission process.

\begin{figure}[t]
	\includegraphics[width=1.0\columnwidth]{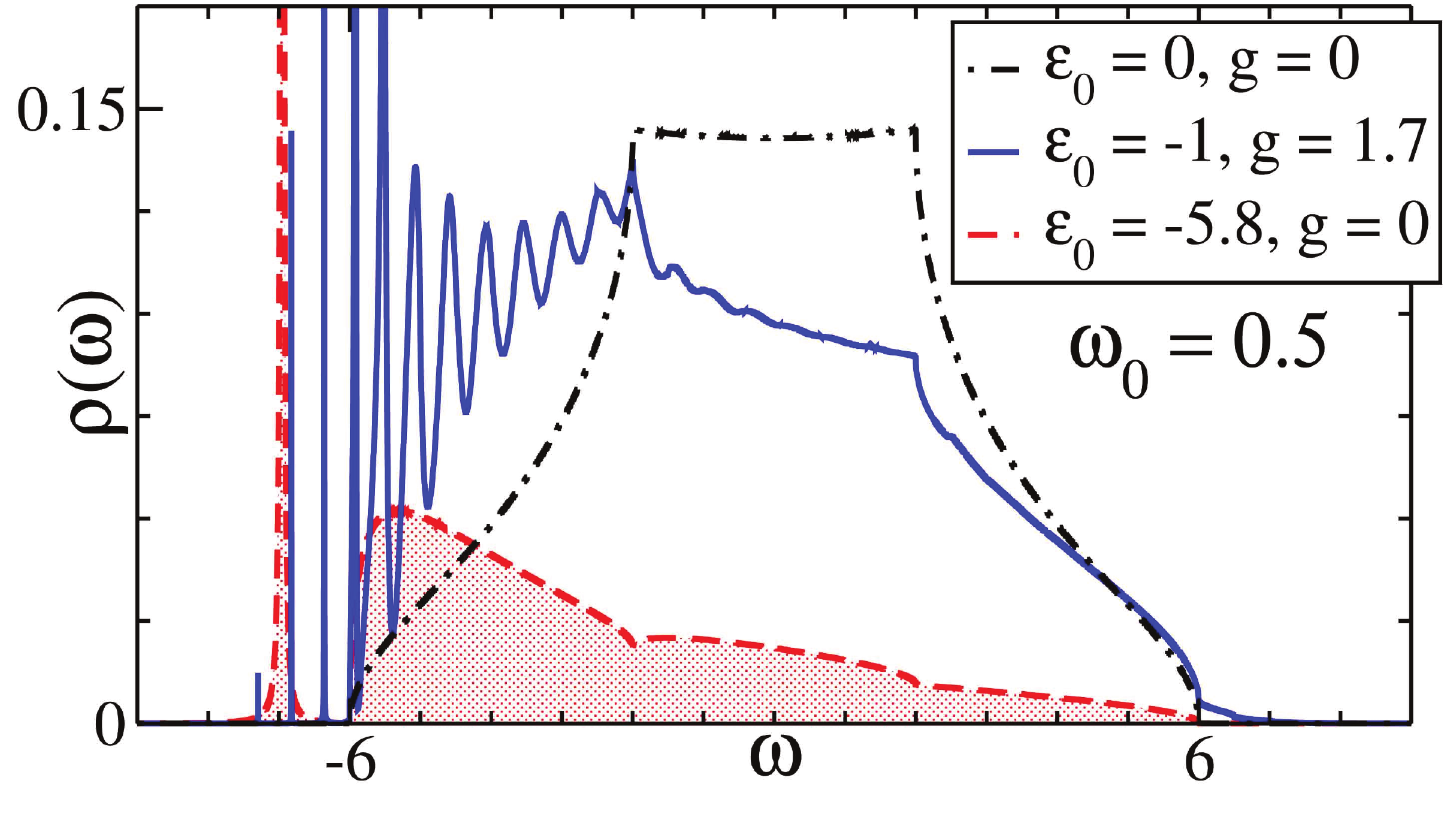}
	\caption{Exact LDOS at the impurity site for different impurity parameters, $\rho(\omega)=-\pi^{-1} \IM G^{0,0}_{\mathbf{l},\mathbf{l}}$.}
	\label{fig03}
\end{figure}

{\it Local properties.} We turn now to a 3D cubic lattice problem, investigating the wide band regime. In all our calculations, we fix the nearest-neighbor hopping, $t=1$, as the unit of energy. We fix the phonon energy, $\omega_0 = 0.5$, as well, unless explicitly stated otherwise, and vary the impurity parameters, $\varepsilon_0$ and $g$. With the exact form of GF in Eq.~(\ref{TM}) known, we evaluate the exact local density of states (LDOS) at the impurity site, $\rho(\omega)=-\pi^{-1} \IM G^{0,0}_{\mathbf{l},\mathbf{l}}$, shown in Fig.~\ref{fig03}. The dot-dashed curve represents the unperturbed LDOS, while the LDOS given by the dashed curve is obtained by introducing the static impurity, $\varepsilon_0=-5.8$, strong enough for a localized bound state to appear below the continuum of delocalized states. Namely, unlike for systems with reduced dimensionality, for $g=0$ in 3D systems, a localized state exists only if the static impurity is sufficiently strong, $|\varepsilon_0|\gtrsim$ 3.96 for the cubic lattice \cite{Ebrahimnejad}. 

The case with a strong EPI, $g=1.7$, $\varepsilon_0=-1$, is shown by the full curve in Fig.~\ref{fig03}, characterized by multiple resonances below the continuum, corresponding to localized states. The lowest resonance involves a large polaronic lattice deformation, i.e., a heavily dressed electron. Consequently, the corresponding electron spectral weight is strongly suppressed. The dressing effect becomes weaker for resonances closer to the continuum, associated with excitations of the polaronic lattice deformation. These excitations are harmonic for deep states and weakly softened in comparison to the bare phonon energy $\omega_0$. As the localized states approach the continuum, some anharmonicity in the excitation energies becomes apparent as well. For $g=1.7$ in Fig.~\ref{fig03}, the phonon nature of excitations at the impurity site is also clearly observed in the part of the LDOS belonging to the delocalized states. Although broadened, being embedded in the continuum of states, the resonances are still well defined, giving rise to the resonant scattering of electrons on the impurity.

{\it Mobility.} The electron scattering is fully described by the $T$-matrix operator, which sums over all scattering events to the infinite order in $\hat V$ \cite{economou}, $\hat G = \hat G^{(0)} + \hat G^{(0)} \hat{ \mathcal{T}}\hat G^{(0)}$. For the problem in Eq.~(\ref{Ham}), $\hat{ \mathcal{T}}$ involves the impurity site only and its matrix elements may simply be read from Eq.~(\ref{TM}), $\mathcal{T}^{\gamma,\alpha}=\Gamma^{\gamma,\alpha}/(1-G^{(0)\alpha,\alpha}_{\textbf{l},\textbf{l}}\Gamma^{\alpha,\alpha})$.

When $\hat{\mathcal{T}}$ is local, as in the present case, the anisotropy of the scattering amplitude,

\begin{equation}\label{amplitudes}
\left\langle \textbf{r},\gamma| \psi \right\rangle =  e^{i\textbf{k}\textbf{r}}\delta_{\gamma,\alpha}+ G^{(0)\gamma,\gamma}_{\mathbf{r},\mathbf{l}} \mathcal{T}^{\gamma,\alpha}\;,
\end{equation}

\noindent is governed only by $G^{(0)\gamma,\gamma}_{\mathbf{r},\mathbf{l}}$. However, since we are mostly interested in the low-frequency electrons close to the bottom of the conduction band, the exact form of $G^{(0)\gamma,\gamma}_{\mathbf{r},\mathbf{l}}$ \cite{callaway,katsura} in Eq.~(\ref{amplitudes}) may be approximated by its isotropic low-frequency form, corresponding to the outgoing s-wave. On the other hand, for high-order diagrams contributing to $\mathcal{T}^{\gamma,\alpha}$ in Eq.~(\ref{amplitudes}) at large $g$ and $\varepsilon_0$, we preserve the exact form of the local propagator $G^{(0)\alpha,\alpha}_{\mathbf{l},\mathbf{l}}$.

\begin{figure}[t]
	\includegraphics[width=1.0\columnwidth]{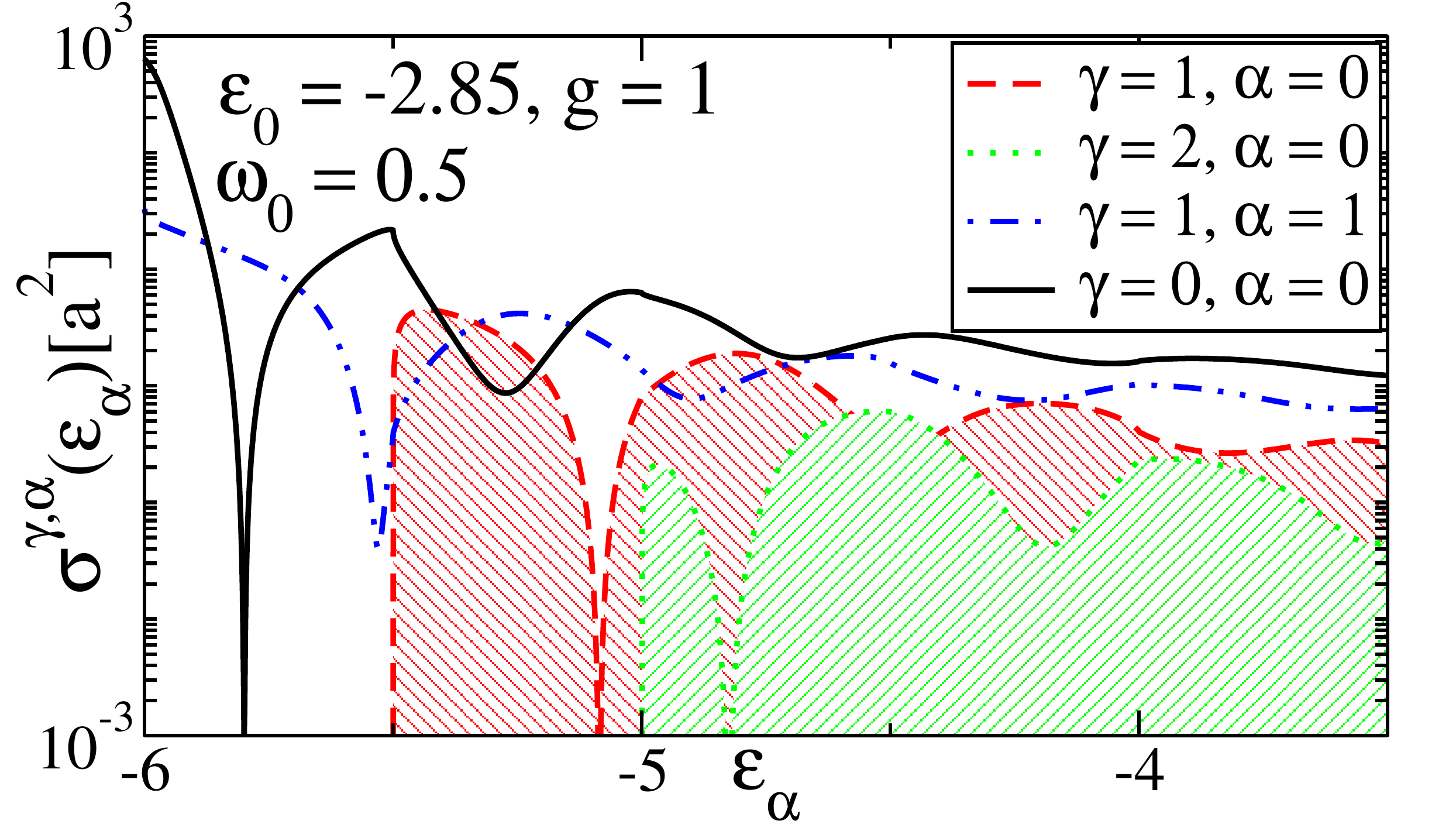}
	\caption{Partial cross-sections (\ref{sigmaag}) for few elastic and inelastic scattering channels as a function of the incoming electron energy $\varepsilon_\alpha$.}
	\label{fig04}
\end{figure}

The partial cross-sections for few elastic and inelastic scattering channels are shown in Fig.~\ref{fig04} as a function of the incoming electron energy $\varepsilon_\alpha$,

\begin{equation}\label{sigmaag}
\sigma^{\gamma,\alpha}(\varepsilon_\alpha)=
\frac{a^2}{4\pi t^2}\sqrt{\frac{\varepsilon_\gamma}{\varepsilon_\alpha}}
|\mathcal{T}^{\gamma,\alpha}(\varepsilon_\alpha+\alpha\omega_0)|^2\;,
\end{equation}

\noindent with $\varepsilon_\gamma=\varepsilon_\alpha+(\alpha-\gamma)\;\omega_0$, the energy of the outgoing electron and $a^2=1$ the area associated to a unit cell. For the nearly resonant choice of parameters in Fig.~\ref{fig04} there are no localized states below the continuum.  The phonon frequency scale, introduced in the LDOS by the EPI, has its strong reflection in Fig.~\ref{fig04}. That is, $\omega_0$ characterizes the energy thresholds for the inelastic scattering involving phonon emission ($\gamma>\alpha$) and governs to a great extent the energy-dependence of all $\sigma^{\gamma,\alpha}$, particularly for low energies $\varepsilon_\alpha$ of the incoming electron, when $\sigma^{0,0}$ in Fig.~\ref{fig04} reaches very large values. Not shown in Fig.~\ref{fig04} are the phonon absorption channels that are related to the phonon emission channels by the time-reversal symmetry, $\mathcal{T}^{\gamma,\alpha}(\omega)=\mathcal{T}^{\alpha,\gamma}(\omega)$, i.e., $\varepsilon_\gamma \sigma^{\gamma,\alpha}(\varepsilon_\alpha)=\varepsilon_\alpha\sigma^{\alpha,\gamma}(\varepsilon_\gamma)$.

For the system in thermal equilibrium, the total cross section $\langle\sigma(\varepsilon_\alpha)\rangle_T$, as a function of the incident electron energy $\varepsilon_{\alpha}$, is obtained simply by averaging over the phonon thermal distribution for initial states ($k_B=1$), 

\begin{equation}\label{sigmatot}
\langle\sigma(\varepsilon_\alpha)\rangle_T=
(1 - e^{-\omega_{0}/T})\sum_{\alpha,\gamma} e^{-\alpha\omega_{0}/T}
\sigma^{\gamma,\alpha}(\varepsilon_{\alpha})\;.
\end{equation}

\noindent Thus, for non-degenerate electrons in semiconductors, the electron mobility may be calculated from \cite{Ziman,Suppl},

\begin{equation}\label{mu}
\mu(T) = \frac{8|e|t}{3\sqrt{\pi}T^{\frac{5}{2}}}\int \varepsilon^{\frac{3}{2}} \;\tau(\varepsilon,T)e^{-\varepsilon/T}\;d\varepsilon\;,
\end{equation}

\noindent where $\tau(\varepsilon,T)$ is the energy- and temperature-dependent electron relaxation time, $\tau^{-1}(\varepsilon_{\alpha}, T) =|v_\alpha|\;n_i\;\langle\sigma(\varepsilon_\alpha)\rangle_T$, with $|v_\alpha|=\sqrt{4t\varepsilon_\alpha}$ representing the electron velocity \cite{Matthiessen}. 

\begin{figure}[t]
	\includegraphics[width=1.0\columnwidth]{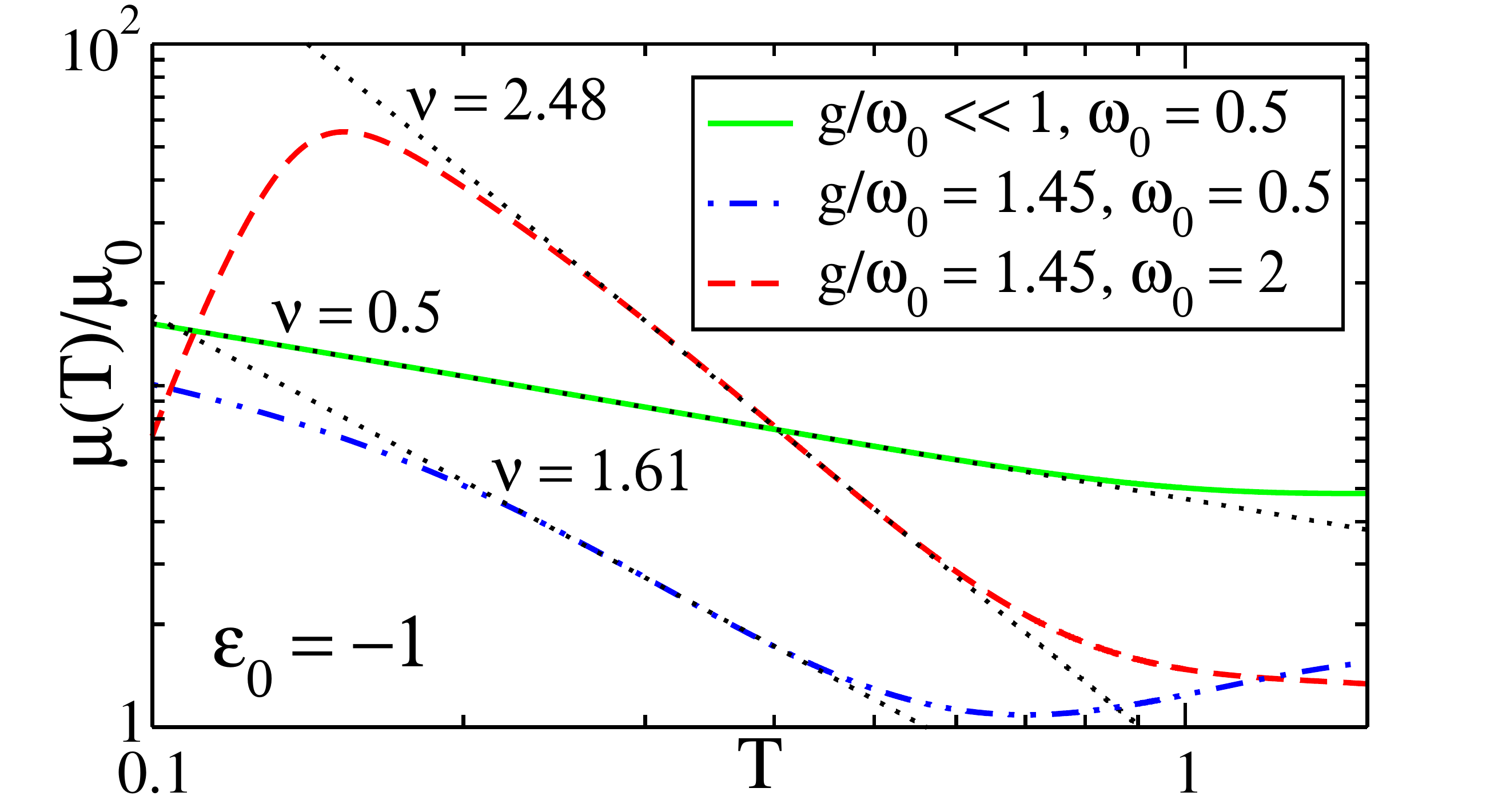}
	\caption{Mobility $\mu(T)$ as a function of temperature for different couplings $g$, shown in the log-log scale.}
	\label{fig05}
\end{figure}

Hereafter, as the reference value for the mobility, we use $\mu_0=\lim_{|\varepsilon_0|\rightarrow\infty }\mu(T=\omega_0)$, when the impurity behaves as a vacancy, with $\mathcal{T}(\omega)=-[G^{(0)}_{\textbf{l},\textbf{l}}(\omega)]^{-1}$. The impurity parameters may achieve very different values in real materials, affecting the mobility over a wide temperature range. However, we focus our attention on the experimentally most relevant temperatures, $T\lesssim\omega_0$, for which the effects of polaronic impurities may be more easily detangled from thermally activated scattering processes on lattice acoustic phonons that frequently dominate on higher temperatures. In particular, for $T<\omega_0$, we find in a broad range of impurity parameters power-law behaviors of $\mu(T)$. For three sets of parameters, $\mu(T)$ is shown in the log-log scale in Fig.~\ref{fig05}. In the weak-coupling limit, $g/\omega_{0}\ll 1$, the power-law $T^{-\nu}$ behavior spans almost over the whole temperature range shown, with $\nu=1/2$ given by the curve slope. This behavior may easily be rationalized by noting that for a weak EPI $\tau(\varepsilon,T)$ is characterized by a nearly constant cross section for low incident energies $\varepsilon_\alpha$. $\nu=1/2$ for weak static (charge-neutral \cite{Sclar}) impurities may be explained in the same way, $\mathcal{T}(\omega)\approx\varepsilon_0$. When frequencies at which $\mathcal{T}(\omega)$ deviates from the constant start to be thermally relevant a weak upturn of $\mu(T)$ in Fig.~\ref{fig05} occurs. As seen from Fig.~\ref{fig05}, for a stronger EPI, $g/\omega_{0}=1.45$, $\mu(T)$ is characterized by a different power-law behavior, $\nu\approx1.61$. For the same ratio $g/\omega_{0}=1.45$, but the larger phonon energy, $\omega_0 = 2$, the low-temperature maximum is followed by the power-law behavior, with $\nu\approx2.48$. The $g/\omega_{0}=1.45$ cases fall in the essentially non-perturbative regime, as it implies contributions  from processes described by many high-order diagrams, with many different scattering channels and partial cross-sections becoming increasingly important upon increasing the temperature \cite{Suppl}.

\begin{figure}[t]
	\includegraphics[width=1.0\columnwidth]{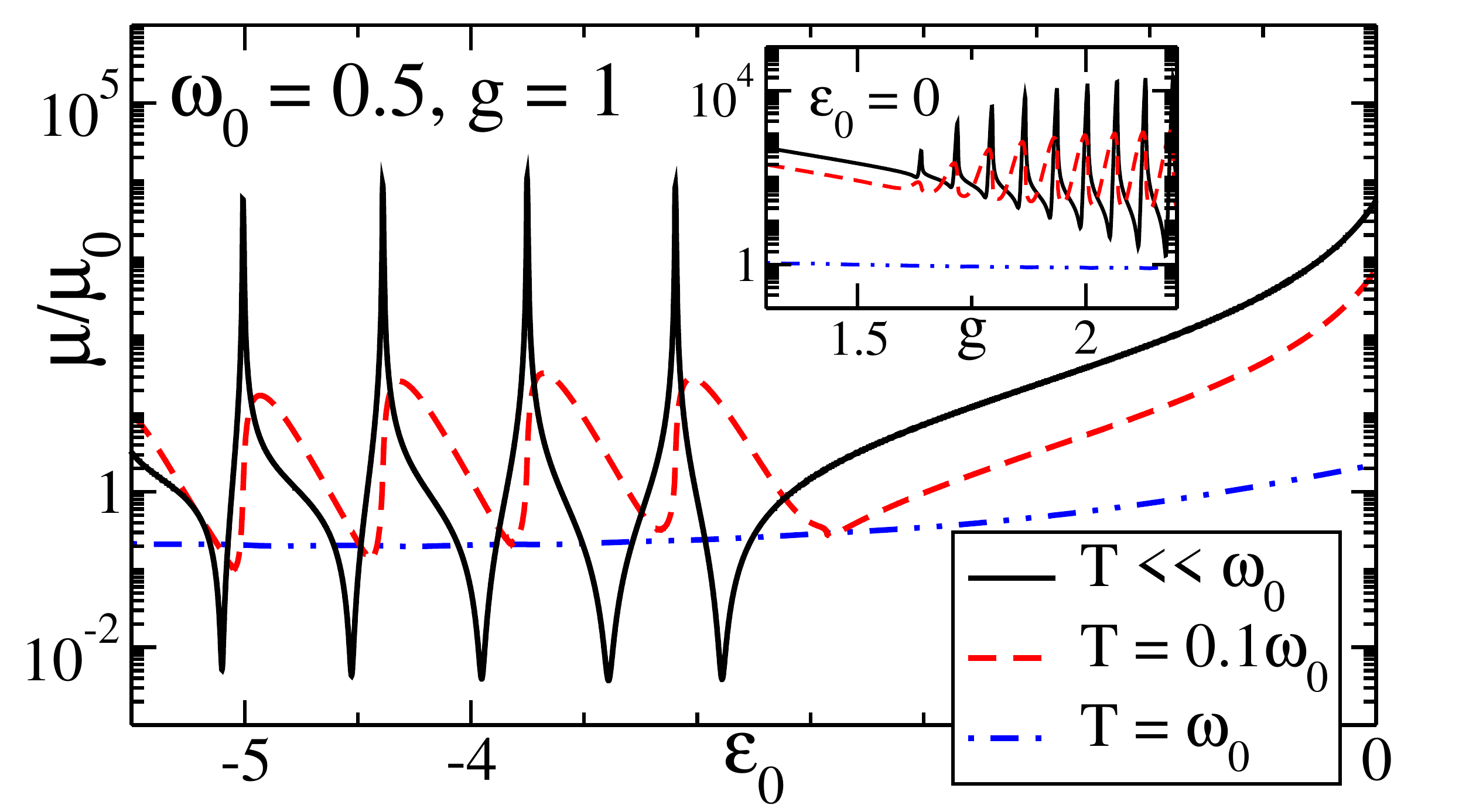}
	\caption{$\mu(T)$, shown in the log scale, as a function of $g$ (inset) and $\varepsilon_0$.}
	\label{fig06}
\end{figure}

The sensitivity of $\mu$ on the impurity parameters and $T$ is investigated in Fig.~\ref{fig06}. When the model parameters satisfy resonant scattering conditions, $\mu$ drops sharply for $T\approx0$, corresponding to a large residual resistivity due to a strong relaxation of electron momenta. In the $T\rightarrow0$ limit, for some parameters in Fig.~\ref{fig06} one observes a fully transparent behavior of polaronic impurities as well. This behavior corresponds to zeros of $\mathcal{T}^{0,0}\propto\Gamma^{0,0}$. In general, all singularities are sensitive to the electron incident energy and get averaged out at elevated $T$. In particular, as $T$ approaches $\omega_0$ in Fig.~\ref{fig06}, thermal averages in Eq.~(\ref{mu}) over the electron and the phonon distributions make $\mu$ a smooth function of impurity parameters.

{\it Conclusions.} We solve exactly the problem of electrons in 3D semiconductor crystals strongly scattered by impurities with local deformation modes. The problem, formulated through the minimal model in Eq.~(\ref{Ham}), is addressed within the Green's function approach, and solved through the continued fraction expansion. The solution allows for the direct interpretation in terms of the diagrammatic expansion to arbitrary order. Valid for arbitrary values of model parameters, system dimensionality and geometry, it allows to explore system's local and transport properties. In particular, the temperature dependence of the electron mobility is shown to generally exhibit the power-law behavior, $\mu(T)\propto T^{-\nu}$. The exponent $\nu$, starting at $\nu=1/2$ for the non-resonant scattering and weak EPIs, changes rapidly upon increasing the EPI. The scattering on polaronic impurities thus imposes as the possible explanation for the broad range of power-law exponents observed experimentally in various organic crystals. \cite{Karl,Xi,Krupskaya,Podzorov}. Our solution readily extends to various system of current interest, including atomically thin crystalline films, where the effects of impurities with internal degrees of freedom can be experimentally accessed through energy-loss spectroscopy, atom-probe tomography and scanning tunneling microscopy \cite{Krivanek,Berthe,Marion}.

{\it Acknowledgements.} J. K., E. T. and O. S. B. acknowledge the support of the Croatian Science Foundation Project IP-2016-06-7258. O. S. B. acknowledges the support by the QuantiXLie Center of Excellence, a project co-financed by the Croatian Government and European Union  through the European Regional Development Fund - the Competitiveness and Cohesion Operational Programme
(Grant KK.01.1.1.01.0004).

\end{document}